# The *s* Process and Beyond

Maria Lugaro,[1,2,3,4] Marco Pignatari,[1,2,5,6] René Reifarth,[7,8] and Michael Wiescher[6,8]

[1]Konkoly Observatory, Research Centre for Astronomy and Earth Sciences (CSFK), Eötvös Loránd Research Network (ELKH), Budapest, Hungary; email: maria.lugaro@csfk.org

[2]Research Centre for Astronomy and Earth Sciences (CSFK), MTA Centre of Excellence, Budapest, Hungary

[3]Institute of Physics and Astronomy, ELTE Eötvös Loránd University, Budapest, Hungary

[4]School of Physics and Astronomy, Monash University, Clayton, Australia

[5]Milne Centre for Astrophysics, University of Hull, Kingston upon Hull, United Kingdom

[6]Joint Institute for Nuclear Astrophysics–Center for the Evolution of the Elements, East Lansing, Michigan, USA

[7]Department of Physics, Goethe-Universität Frankfurt, Frankfurt am Main, Germany

[8]Department of Physics and Astronomy, University of Notre Dame, Notre Dame, Indiana, USA

## Abstract

Neutron captures produce the vast majority of abundances of elements heavier than iron in the Universe. Beyond the classical slow (*s*) and rapid (*r*) processes, there is observational evidence for neutron-capture processes that operate at neutron densities in between, at different distances from the valley of $\beta$ stability. Here, we review the main properties of the *s* process within the general context of neutron-capture processes and the nuclear physics input required to investigate it. We describe massive stars and asymptotic giant branch stars as the *s*-process astrophysical sites and discuss the related physical uncertainties. We also present current observational evidence for the *s* process and beyond, which ranges from stellar spectroscopic observations to laboratory analysis of meteorites.

## Contents

## 1. INTRODUCTION

The light elements up to iron are produced in stars by proton and α captures (H, He, C, Ne, O, and Si burning), and the iron peak is mostly the result of nuclear statistical equilibrium. The elements heavier than iron are instead made by neutron captures (1, 2) with only a small number of proton-rich nuclei created by photodisintegration (3). The reason is that capture of charged particles by heavy elements is suppressed by their strong Coulomb barrier. Captures of neutrons, by contrast, are easy to achieve even at room temperature if free neutrons are available. As neutrons decay into protons, in nature they are typically rare. Therefore, an intrinsic problem of investigating neutron captures is the production of the neutrons themselves via nuclear reactions in stars and their behavior in exotic astrophysical objects, such as neutron star mergers.

Similar to the elements lighter than Fe, the solar abundance distribution of the heavier elements presents peaks corresponding to the presence of magic nuclei, mostly in terms of neutrons but also in terms of protons (as in the case of Sn; see **Figure 1**). Such features allowed early research to introduce the main mechanisms of neutron captures in nature: the slow ($s$) and rapid ($r$) processes (1).[1] These two processes are located at the two extremes of the possible neutron density ($N_n$) range (i.e., ~$10^6$ and >$10^{20}$ cm$^{-3}$) and were originally introduced independently from the source of neutrons and the astrophysical site. The $s$ and $r$ processes were defined by neutron-capture timescales that were slower or more rapid, respectively, than the β-decay timescale of unstable nuclei. These processes create the vast majority of abundances of elements heavier than iron in the Solar System.

A first $s$-process site was identified in the 1950s thanks to the observations of peculiar giant stars with enhanced abundances of $s$-process elements [e.g.,Ba (1,5)] and the presence of the radioactive

**Neutron density ($N_n$):** number of neutrons per cubic centimeter available during a neutron flux; it is a function of both time and location

---

[1] Throughout this review, we use the traditional notation introduced in Reference 1 of writing the terms $s$ process and $r$ process without hyphenation. The terms are hyphenated when used adjectivally (e.g., $s$-process elements).

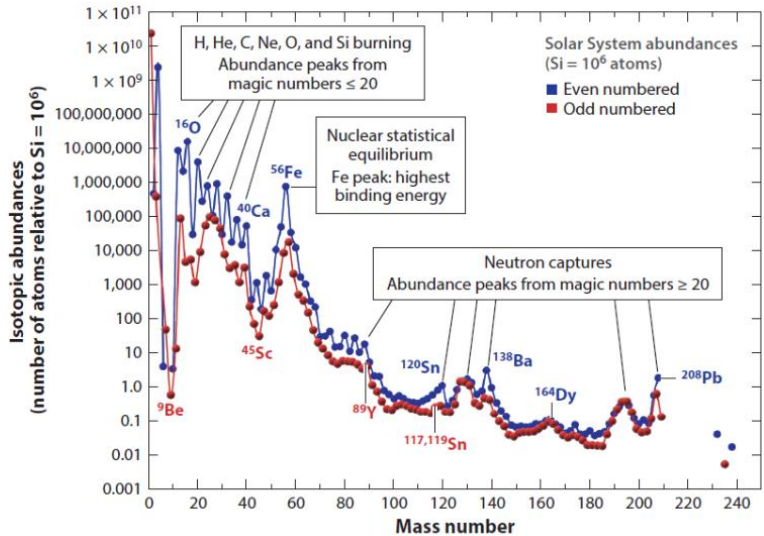

**Figure 1**

Isotopic abundances (by number of atoms and normalized to Si = $10^6$) of the isotopes in the Solar System measured from meteorites as a function of their atomic mass. The elements heavier than iron make up more than two-thirds of all the types of atomic nuclei in the Universe, but their abundances are roughly five orders of magnitude lower than that of Fe. In the Earth's upper continental crust, by contrast, their relative abundances are higher by two orders of magnitude. The nuclear structure features that produce the peaks in the distribution are indicated along with their corresponding main production processes. Figure adapted with permission from Reference 4; copyright Oxford University Press.

element Tc (6), which could be explained only by neutron captures occurring within the observed stars. For the $r$ process, instead, the first observational evidence of its occurrence in an astronomical object (a neutron star merger event) was reported much later—in 2017, thanks to the famous coincidence of the LIGO GW170817 neutron star merger event with the AT 2017gfo kilonova and the GRB 170817A short γ-ray burst (7, 8). Since the introduction of the $s$ and $r$ processes, other neutron-capture processes, such as the intermediate ($i$) and neutron ($n$) processes (the latter also referred to as neutron bursts), have also been defined and investigated (**Table 1**). While the $i$ and $n$ processes, as well as the astrophysical sites mentioned in **Table 1**, are significant topics themselves, we do not cover them in detail in this review, but we will introduce them in relation to the $s$ process and discuss them when relevant for specific observational constraints.

## 2. NEUTRON-CAPTURE PROCESSES BETWEEN THE $s$ AND $r$ PROCESSES

A huge variety of possible neutron fluxes may exist in nature with neutron densities between those of the $s$ and $r$ processes. While the $i$ and $n$ processes cover most of the possible neutron density ranges (**Table 1**), they have been investigated so far mostly in reference to specific conditions related to the stellar site where they have been assumed and/or predicted to occur. The $i$ process was introduced more than four decades ago on the basis of theoretical calculations of giant stars (9). While there is no clear indication of it when considering the Solar System distribution of **Figure 1**,

www.annualreviews.org • The s Process and Beyond    317

**Table 1** The slow, intermediate, neutron, and rapid processes

| Name(s) | $N_n$ (cm$^{-3}$) | Neutron source(s) | Astrophysical site(s) | Reference(s)[a] |
|---|---|---|---|---|
| Slow (s) | $10^6$–$10^{11}$ | $^{13}$C($\alpha$,n)$^{16}$O | AGB[b] stars | 25–27 |
| | | $^{22}$Ne($\alpha$,n)$^{25}$Mg | Massive stars[c] | 28–30 |
| Intermediate (i) | $10^{12}$–$10^{15}$ | $^{13}$C($\alpha$,n)$^{16}$O | Post-AGB stars[d] | 10, 11, 17 |
| | | | Low-$Z$[e] AGB stars | 12, 13, 15, 31 |
| | | | Super-AGB stars[f] | 32 |
| | | | Accreting white dwarfs | 11, 33, 34 |
| | | | Massive stars[c] | 35–37 |
| Neutron (n) (also called neutron burst) | $10^{18}$–$10^{20}$ | $^{22}$Ne($\alpha$,n)$^{25}$Mg | He shell of CCSNe[g] | 22, 23 |
| Rapid (r) | >$10^{20}$ | — | Compact mergers[h] | 38–40 |
| | | | Special CCSNe[i] | |

[a] The Reference(s) column includes examples of reviews and articles for each process; for the *i* and *n* processes, reviews are not available yet.
[b] The asymptotic giant branch (AGB) is the final evolutionary phase of stars with masses below ∼8 $M_\odot$, which burn only H and He in their cores. The *s* process in AGB stars is discussed in detail in Section 5.
[c] Massive stars have masses above ∼10 $M_\odot$ and end their lives as core-collapse supernovae (CCSNe) after their cores experience H to Si burning. The *s* process in massive stars is discussed in more detail in Section 4 and at length in Reference 41. [d] The post-AGB is the evolutionary phase of low-mass stars between the AGB and the white dwarf.
[e] Low-$Z$ stars have low metal content—roughly below 1/10$^4$ of the solar $Z \simeq 0.016$.
[f] Super-AGB stars have masses in the range 8–10 $M_\odot$ and ignite carbon in their core before entering the AGB phase. They leave an Ne-O white dwarf rather than a C-O white dwarf as their lower-mass counterparts do.
[g] A CCSN results from the collapse of the iron core of a massive star.
[h] Compact mergers include neutron star–neutron star and neutron star–black hole mergers.
[i] Special CCSNe are those that lead, for example, to the formation of collapsars and magnetars.

**Thermal pulse (TP):** event in which energy is released within a short timescale by He burning at the base of an He-rich shell, which causes the whole shell to become convective

**Metallicity:** sum of the abundances of all elements except H and He, which are therefore all "metals"; metallicity is indicated by the symbol $Z$ (not to be confused with the atomic number, $Z$) the *i* process has gained recent popularity due to potential detection of its signature in stars.[2] These range from the Sakurai's Object, whose fast evolution is a result of a very late thermal pulse (TP) (see Section 5) (10, 11), to the fraction of carbon-enhanced metal-poor (CEMP) stars showing excesses of both Ba and Eu (an *s*- and an *r*-process element, respectively), now named CEMP-*i* (12–15), and a number of post-AGB (asymptotic giant branch) stars of metallicity roughly 1/20 solar (16, 17).

The neutron burst, or *n* process (20), has been considered to explain the isotopic features of the Xe-H component present in meteoric stardust diamonds showing enrichment in the heavy isotopes of Xe (at masses of 134 and 136) different from those expected from the *r* process (21). It has also been investigated to explain the Mo and Zr isotopic abundances measured in meteoritic stardust silicon carbide (SiC) grains from core-collapse supernovae (22, 23). The neutron burst is driven by a small number of free neutrons in the He shell during the passage of the supernova shock on a timescale of roughly 1 s (24). Therefore, it does not produce significant abundances of elements heavier than iron, but it is mostly a local process responsible for shifting the initial abundances of each element toward those of its neutron-rich isotopes.

The *s* process stands out from the other three neutron-capture processes because its path of neutron captures proceeds mostly along the stable nuclei on the valley of β stability (**Figure 2**). For the other processes, the path of neutron captures proceeds mostly along unstable nuclei to the right of the valley of β stability. Therefore, the *s*-process final abundances are determined mostly during the neutron flux, while for the other processes a last step is needed to derive the final

---

[2] Potential *i*-process signatures are also observed in the elements lighter than Fe in some types of stardust grains (18, 19).

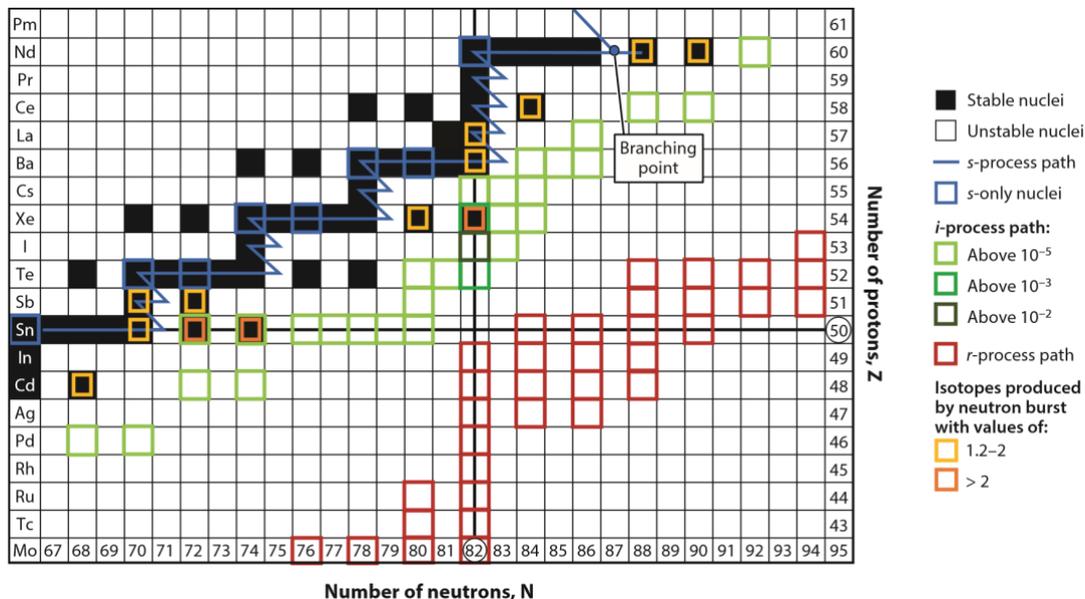

**Figure 2**

The four different neutron-capture processes illustrated on a section of the nuclide chart. The black and white boxes are stable and unstable nuclei, respectively. The magic numbers of protons (Z = 50) and neutrons (N = 82) are indicated by the heavier solid black lines. The stable isotopes located on the proton-rich side (i.e., on the left) of the valley of β stability cannot typically be produced by neutron captures and are referred to as *p*-only isotopes. The blue line represents the *s*-process path following the valley of β stability. As an example, the branching point at $^{147}$Nd, with a half-life of roughly 11 years, is indicated on the top right. Blue boxes represent *s*-only nuclei—that is, those isotopes that are shielded from the decay of *i*- and *r*-process products by a stable isotope of the same mass belonging to a lighter element. The green boxes represent a typical *i*-process path during a neutron flux with a neutron density of $10^{15}$ cm$^{-3}$. Light, medium, and dark green boxes are isotopes with mass fractions above roughly $10^{-5}$, $10^{-3}$, and $10^{-2}$ ($^{135}$I only), respectively (from figure 3.5 of Reference 42). The red boxes (data from an image from Reference 43, courtesy of K.-L. Kratz and H. Schatz) represent the *r*-process path using the location of the *r*-process waiting points (according to the ETFSI-Q mass model), where an equilibrium between neutron-capture rates (n,γ) and photodisintegration (γ,n) is reached and the flow has to wait for β decays to proceed. The final results of the *i* and *r* processes are obtained from the decay of the abundances on the paths shown here. The orange boxes represent isotopes produced by the neutron burst already after the final decay, with lighter and darker orange boxes representing isotopes with values between 1.2 and 2 and greater than 2, respectively (data from Reference 21, table 1, last column).

predictions: After the neutron flux is extinguished, all the abundances must be decayed into stable isotopes. A similarity among the *s*, *i*, and *r* processes is related to the presence on their paths of nuclei with magic numbers. These nuclei determine the final abundance distribution because their neutron-capture cross sections are orders of magnitude lower than those of the other heavy nuclei. For the process of the neutron burst, magic nuclei are not as relevant because the process is mostly local. The magic numbers are well known to have shaped the solar abundance distribution, where the abundance peaks related to the *r* and *s* processes are clearly visible (**Figure 1**). The difference between the two types of peaks is that the *s*-process peaks represent the stable magic isotopes $^{88}$Sr, $^{89}$Y, and $^{90}$Zr (the first *s*-process peak, with N = 50); $^{138}$Ba, $^{139}$La, and $^{140}$Ce (the second *s*-process peak, with N = 82); and $^{208}$Pb (the third *s*-process peak, with N = 126). For the *r* process, the peaks instead correspond to the result of the decay of unstable magic isotopes. Therefore, they are shifted to lower masses relative to the *s*-process peaks (i.e., roughly at Se, Xe, and Pt; see **Figure 1**). A similar effect is present when modeling the *i* process; for example, in this case the magic stable isotope $^{86}$Kr and the magic unstable isotope $^{135}$I have the highest abundances (**Figure 2**). After decay, the latter results in a high abundance of Ba (relative to, e.g., La) and specifically of $^{135}$Ba, which is a fingerprint of the *i* process (44).

During the *s* process, nuclei with a magic number of neutrons generate bottlenecks on the path of neutron captures and accumulate until their abundance becomes high enough to generate a significant probability of capturing a neutron. The first bottleneck at Sr, Y, and Zr needs to be overcome before the second bottleneck at Ba, La, and Ce can be reached (see also Reference 45, figure 2); therefore, an increasingly higher time-integrated neutron flux [neutron exposure ($\tau$)] is required to reach the second and then the third peak of the *s* process at Pb. Here, the *s* process terminates because, beyond Pb and Bi, unstable nuclei predominantly $\alpha$ decay rather than $\beta$ decay. The atomic number always decreases instead of increasing, and the flux recycles back to stable Pb and Bi. Therefore, all elements beyond Pb and Bi are produced by the *r* process, although potential contributions from the *i* process are also possible and are still under investigation.

Between magic numbers, the *s* process behaves in an extremely predictable way. Once a bottleneck is bypassed, the neutron-capture flow can proceed to fill the abundances of all the nuclei up to the next bottleneck, and the abundances between the peaks reach steady-state equilibrium. Therefore, their neutron-capture cross sections determine their relative abundances according to the following rule of thumb: $N_A \langle \sigma v \rangle_A \simeq$ constant, where $N_A$ is the abundance of an isotope of mass A during the *s* process and $\langle \sigma v \rangle_A$ is the average of the neutron-capture cross section $\sigma_A$ over the distribution of the velocities v. The astrophysical site, and specifically the temperature at which the neutrons are released, plays a role when the value of $\sigma_A$ depends on the temperature and, therefore, the velocity. This is not the case for most isotopes (as usually $\sigma_A \approx 1/v$), but it can occur in some specific cases and should be considered when using the simple rule above. A significant consequence of the establishment of steady-state equilibrium during the *s* process is that some abundance patterns cannot be produced by the *s* process. If observed, these necessarily require the activation of processes beyond the *s* process. For example, in the *s*-process framework, high abundances of, for instance, Hf and W located between the second and third peaks cannot be produced without a correlated high abundance of the third peak at Pb. This feature indicates the occurrence of the *i* process in CEMP and Pb-poor post-AGB stars (16, 17). By operating farther away from the valley of $\beta$ stability, the *i* process allows the neutron-capture flow to produce unstable nuclei that can decay into, for instance, Hf and W without increasing the abundance of Pb. There is also a fraction of Ba stars that show abundances of the elements just beyond the first and second peaks higher than *s*-process model predictions (see Section 6.1). This feature is impossible to explain by the steady-state equilibrium pattern of the *s* process.

Unstable isotopes with half-lives of at least a few days can capture neutrons even during the *s* process. They become so-called branching points, at which the flow splits between two different paths, one of which moves slightly away from the valley of $\beta$ stability. An example is indicated by $^{147}$Nd in **Figure 2**, where neutron captures compete against decay to produce $^{148}$Nd. The appendix of Reference 46 reports a full list of the roughly 60 branching points potentially activated along the *s*-process path, together with brief descriptions of the characteristics and implications of each of them. There are many hidden complexities in the calculation of the neutron-capture flux at branching points. Decay rates may depend on the temperature (in which case the branching point may be used as a thermometer) and the density, and the result is controlled by the neutron density, which depends on the characteristics of the astrophysical site and the rate of the neutron-source reactions. Furthermore, the neutron density varies with time, and the final abundances of isotopes affected by branching points are determined by the whole temporal evolution of the temperature and the neutron density. All these properties carry uncertainties, and therefore the study of branching points is particularly elaborate. Branching points are not responsible for modifications of the elemental *s*-process abundance pattern except for the case of the Rb abundance, which

**Neutron flux:** refers generically to an event when free neutrons are available, and specifically to the neutron density multiplied by the velocity: $N_n \times v$

**Neutron exposure ($\tau$):** integral of the neutron flux over a given time interval; the total neutron exposure is calculated over the whole interval of the availability of free neutrons, up to the time when the neutron flux is extinguished depends on the activation of the branching point at $^{86}$Rb feeding the magic $^{87}$Rb (47,48). Detailed constraints can be derived from the solar isotopic abundances, especially from the relative abundances of s-only isotopes affected by branching points (49)—for instance, the $^{134}$Ba/$^{136}$Ba ratio (**Figure 2**) affected by the branching point at $^{133}$Cs. Data from laboratory analyses of meteoritic stardust are also provided as isotopic abundances, which carry the imprint of branching points (Section 6.2). Furthermore, some branching points are crucial to disentangle the s process from the photodisintegration (γ process) contribution to the solar abundances, in particular for some p-only (e.g., $^{94}$Mo) and s-only isotopes (e.g., $^{128}$Xe) (50).

## 3. NUCLEAR INPUT FOR THE s PROCESS

### 3.1. Neutron-Capture and β-Decay Rates

To solve an s-process network, the energy-dependent neutron-capture cross sections of the stable isotopes are needed as well as those of the unstable isotopes that represent branching points along the main neutron-capture path. The i and n processes share similar needs, although more information on unstable isotopes is required. A popular source of such data is the KADoNiS compilation, which is also included in the JINA REACLIB database (**https://reaclib.jinaweb.org**). The ASTRAL database (**https://exp-astro.de/astral/**) is also currently coming online.

While nuclei in the laboratory usually exist in the ground state, high temperatures in stars result in a significant population of higher-lying states (51). Therefore, the relative contribution of each of these states (52), or a stellar enhancement factor, needs also to be included in the rates used in stellar models. The temperature and density dependence of β-decay and electron-capture rates are also dependent on the population of higher-lying states and necessary to model the behavior of the branching points. While neutron-capture cross sections during the s process in AGB stars typically have a local effect, it is still crucial to know them with a precision within 10%, for example, to compare the model predictions with the high-precision data from stardust grains (Section 6.2). In massive stars, instead, the uncertainties of neutron-capture rates along the s-process path between iron and strontium may be propagated to all the heavier s-process product (53–55), and the precision needs to be better than 20%. Furthermore, several light isotopes behave as neutron poisons, consuming a significant fraction of neutrons. In this case, the uncertainties are propagated to the neutron exposure and over the whole abundance distribution. Famous examples of light neutron poisons are the $^{22}$Ne$(n,\gamma)^{23}$Ne (56, 57), $^{25}$Mg$(n,\gamma)^{26}$Mg (28, 55), and $^{16}$O$(n,\gamma)^{17}$O reactions—the last in particular for low-metallicity stars (58).

The most general approach to determine energy-dependent neutron-capture cross sections is the time-of-flight (TOF) technique (59), as used, for example, in the n_TOF experiment at CERN. Neutrons over a wide energy range are produced in short, intense pulses. The sample material is typically placed from several up to hundreds of meters away and (partly) surrounded by γ-ray or particle detectors (60–62). While this technique is, in principle, applicable to stable and unstable isotopes, the vast majority of measurements have been so far performed on stable samples. The reason is that a very limited amount of radioactive material can be placed close to the necessary γ detectors. Since the radioactivity of a given number of atoms increases with short half-lives, the TOF technique is typically restricted to nuclei with half-lives longer than a few years (63–65). Aside from the sensitivity of the detectors, the sample production is typically a challenge as well because rather large amounts of sample material and enrichment levels higher than approximately 10% are required. The main advantage of the TOF method is that the measured energy-dependent cross sections can easily be converted into Maxwellian-averaged cross sections (MACS) with little additional information required.

The activation method is applicable if the reaction product is not abundant in the sample. This is the case if the reaction product is unstable. During irradiation with neutrons, the reaction product is produced and decays on the timescale of its half-life time (66). If the half-life is not too long, the radioactivity of the product can be used to determine the number of produced nuclei using particle detection techniques. If the half-life is longer than approximately 100 years, the presence of the reaction product itself can be detected using accelerator mass spectrometry (67, 68). Despite the higher sensitivity, activation measurements also are typically restricted to isotopes with half-lives longer than a few years. Since the detection of the products is isotope specific, the activation technique is very sensitive and the sample requirements are much more relaxed in the activation than in the TOF method. In fact, experiments have been successfully performed with submicrogram material embedded in hundreds of milligrams of material (69,70). The main disadvantage, however, is that only spectrum averaged cross sections can be determined. To determine MACS, energy-dependent cross sections need to be derived first, and such derivations require additional theoretical support. To keep these additional assumptions and the resulting systematic uncertainties small, the neutron spectrum during the activation should resemble the stellar neutron energy distribution as closely as possible. Most activation measurements for s-process studies performed to date rely on neutrons from the $^7$Li$(p,n)^7$Be reaction, whose spectrum closely resembles a neutron energy distribution corresponding to the Maxwellian energy distribution at $kT =$ 25 keV—that is, 290 MK (71). Higher temperatures can be reached with the same neutron source via superposition of cross sections that result from different spectra (66). Reaching lower temperatures requires either different neutron sources (72) or special sample geometries, like rings instead of disks.

Finally, measurements in inverse kinematics, where the light reaction partner is at rest while the heavier ion is accelerated, have helped to overcome the half-life limitation for charged-particle-induced reactions (73). In principle, the same approach is possible for neutron-induced reactions. A combination of an ion storage ring and a reactor or spallation source would allow direct measurements on radioactive isotopes down to half-lives of minutes or even less (74,75; see **Figure 3**).

In addition, effective β-decay rates are needed to build an s-process network. To date, such data have been provided mostly by the NETGEN tool (**http://www.astro.ulb.ac.be/Netgen/**) and the compilation by Takahashi & Yokoi (76). The temperature dependence of the half-lives of the branching point isotopes $^{182}$Hf and $^{134}$Cs has recently been reevaluated (77–79). Two important mechanisms play a role in the dependence of effective β-decay rates on the stellar environment. First, as mentioned above, the temperature of the stellar plasma affects the excitation of nuclear states (51). If these states have a different β-decay lifetime, the effective β-decay lifetime of the nucleus may change with the temperature. Most of these higher-lying states are short-lived, and currently the only feasible experimental approach to investigate such rates is through charge-exchange reactions like $(p, n)$ or $(^3$He,$t)$ supported by nuclear theory (78–80). Temperature and electron density also change the population of atomic (electronic) states. This effect is called bound state decay and can dramatically change the half-life of β– decay, since the electron can be emitted into an otherwise filled orbit of the atom. Electron-capture rates are also affected by the electron density and temperature since the likelihood of a bound electron in the vicinity of the nucleus decreases with temperature, whereas the likelihood of a free electron close to the nucleus increases with electron density (76). The effect of increasing ionization can be investigated with ion storage rings. Fully or partly stripped ions can be stored and observed for days, enabling the measurement of half-lives of interest for stellar nucleosynthesis (81).

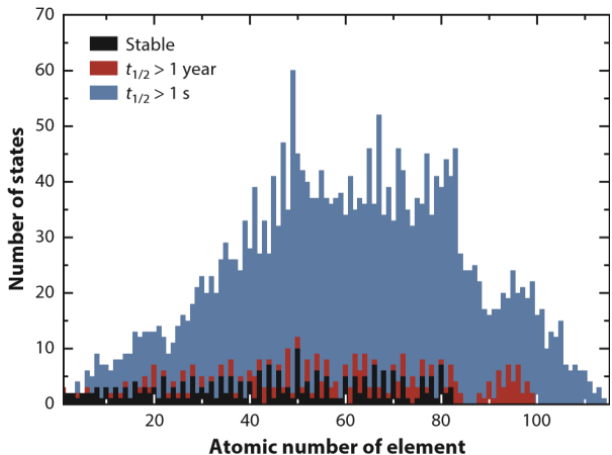

**Figure 3**

Number of the nuclear states of each element for which neutron-induced reactions can be directly investigated. States that are stable or have half-lives longer than 1 year can usually be investigated at current facilities with time-of-flight or activation techniques. While not yet assembled, the combination of a neutron target (e.g., a spallation neutron source) and an ion storage ring with currently available technology would allow the investigation of isotopes with half-lives greater than 1 min. Possible future improvements might even allow the investigation of isotopes with half-lives as short as 1 s, covering the realm of the *s*, *i*, and *n* processes as well as most of the isotopes produced during the freeze-out phase of the *r* process.

### 3.2. Neutron-Source Reactions

When a neutron-capture network is coupled to stellar evolution, the production of neutrons needs to be followed, and the rates of the neutron-source reactions are required as well as those of any other reaction that may affect the production and abundance of the neutron-source nuclei themselves. Determining the reaction rates of all these processes has been a major goal for the experimental nuclear astrophysics community for many decades. The sources of neutrons for the *s*, *i*, and *n* processes are the $^{13}$C($\alpha$,n)$^{16}$O and $^{22}$Ne($\alpha$,n)$^{25}$Mg reactions (**Table 1**). In AGB stars, the neutron flux generated by the $^{13}$C($\alpha$,n)$^{16}$O reaction depends on the amount of $^{13}$C produced by the $^{12}$C($p$,$\gamma$)$^{13}$N($\beta^+$)$^{13}$C proton-capture chain activated within an He/C-rich radiative region that experienced partial mixing (Section 5.1). Also, in the *i* process, mixing of protons into an He-rich region is invoked to produce the $^{13}$C neutron source. However, this is usually through an ingestion into a hotter He-burning convective region (e.g., into the convective region produced by TPs in low-Z AGB stars and post-AGB stars). Therefore, a role is also played by the interplay between the rates of the proton-capture and neutron-source reactions and the dynamical mixing process that transports the unstable $^{13}$N into different temperature zones prior to its decay to $^{13}$C (10). Such complex conditions are becoming accessible via modern multidimensional hydrodynamics simulations, which can inform the classical one-dimensional stellar models (82).

Recent studies of the low-energy range of the $^{13}$C($\alpha$,n)$^{16}$O reaction rate at the deep underground accelerator facilities of LUNA in Italy (83) and JUNA in China (84) have removed most of the uncertainties in the extrapolation of the previous higher-energy data [the NACRE II compilation (85)]. The low-energy data match well the prediction of a recent R-matrix analysis (86) that describes the low-energy cross section in terms of a pronounced interference between the low-energy tails of broad resonances and the unbound extensions of subthreshold states tailing into the effective energy range. Based on the new experimental data and a more sophisticated multichannel R-matrix analysis (87), the new reaction rate is roughly 20% lower than the rate listed in the NACRE II compilation. The overall uncertainty is presently at the level of 20%, and efforts are underway to reduce this value using a new Monte Carlo–based, Bayesian error analysis of the cross-section data (88).

The rate of the $^{22}$Ne($\alpha,n$)$^{25}$Mg reaction carries more substantial uncertainties than that of the $^{13}$C($\alpha,n$)$^{16}$O reaction. This neutron source becomes available in He-burning conditions both in the core of massive stars and in the He-rich shell of AGB stars. It also plays a significant role as a neutron source for the $n$ process when the supernova shock front traverses the He shell of the supernova progenitor star (20). The $^{22}$Ne abundance is orders of magnitude higher than initially in the stellar material because the CNO abundances are converted into $^{14}$N during the H-burning phase and then to $^{22}$Ne via the $^{14}$N($\alpha,\gamma$)$^{18}$F($\beta^+\nu$)$^{18}$O($\alpha,\gamma$)$^{22}$Ne chain during the early phase of He burning. Alternative nuclear reaction paths, such as $^{18}$O($\alpha,n$)$^{21}$Ne($\alpha,n$)$^{24}$Mg (89), can modify the abundance of $^{22}$Ne and therefore the strength of this neutron source. The final amount of $^{22}$Ne typically corresponds to the initial metallicity of the star; for instance, it is above 1% in mass for stars of metallicity around solar. When the temperature is high enough to activate the $^{22}$Ne($\alpha,n$)$^{25}$Mg reaction, $^{22}$Ne is partially consumed and the $s$ process is activated (90, 91). In massive stars, this neutron source is activated also in the C-burning shell at typical temperatures around $10^9$ K, generating an additional $s$-process neutron flux. This is thanks to the production of $\alpha$ particles by C fusion via $^{12}$C($^{12}$C,$\alpha$)$^{20}$Ne and $^{12}$C($^{12}$C,$p$)$^{23}$Na($p,\alpha$)$^{20}$Ne (92–94). Additional reactions affecting the activation of the $^{22}$Ne source in the C-burning shell are $^{12}$C($\alpha,\gamma$)$^{16}$O (29) and the $\alpha$-capture rates on $^{17}$O at low metallicity (89, 96, 97) (see Section 4).

The negative $Q$ value of $^{22}$Ne($\alpha,n$)$^{25}$Mg limits neutron production to high temperature only (roughly >300 MK). At these temperatures, the $^{22}$Ne($\alpha,\gamma$)$^{26}$Mg reaction may also deplete the abundance of $^{22}$Ne during both shell He burning in AGB stars and core He burning in massive stars. Therefore, the efficiency of the $^{22}$Ne($\alpha,n$)$^{25}$Mg depends also on the strength of the competing $^{22}$Ne($\alpha,\gamma$)$^{26}$Mg reaction (98). For example, the $s$-process yields from massive stars with initial solar composition carry an uncertainty of up to a factor of three only due to the $^{22}$Ne($\alpha,n$)$^{25}$Mg uncertainty and up to a factor of 10 when considering the combined impact of uncertainties in both the $\alpha$-capture channels (55, 98–100), as shown in **Figure 4**. During shell C burning in massive stars the main competitor is instead $^{22}$Ne($p,\gamma$)$^{23}$Na (55), which is well measured at the relevant temperatures (101).

A recent analysis suggests that both $^{21}$Ne+$\alpha$ reactions are dominated by the contributions of a pronounced $\alpha$-cluster resonance at a laboratory energy of 830 keV (102). The $^{22}$Ne($\alpha,\gamma$)$^{26}$Mg reaction may also have additional resonant contributions from neutron bound states below 650 keV (99). For the $^{22}$Ne($\alpha,n$)$^{25}$Mg reaction, the strength of the 830-keV resonance is under debate due to discrepancies between direct measurements (103–105) and indirect reaction studies of resonance (106, 107). For lower energy resonances in $^{22}$Ne($\alpha,\gamma$)$^{26}$Mg, a new study (108) confirmed earlier results (109) but was not yet successful in observing the predicted lower energy resonances (99, 110). New experiments in deep underground accelerator laboratories are running to determine the associated resonance strengths (111).

## 4. THE *s* PROCESS IN MASSIVE STARS

Massive stars ($M \gtrsim 10 M_\odot$) were identified as a source of the $s$-process elements by early stellar computational simulations (112) with $^{22}$Ne($\alpha,n$)$^{25}$Mg as the dominant source of neutrons activated mostly during core He burning and shell C burning (29, 55, 93). Classically, the $s$ process in massive stars was identified as the main producer of the weak $s$-process component in the Solar System—that is, the $s$-process abundances between iron and the first-peak elements Sr, Y, and Zr (113) (**Figure 4**). Indirect observations and theoretical nucleosynthesis calculations have

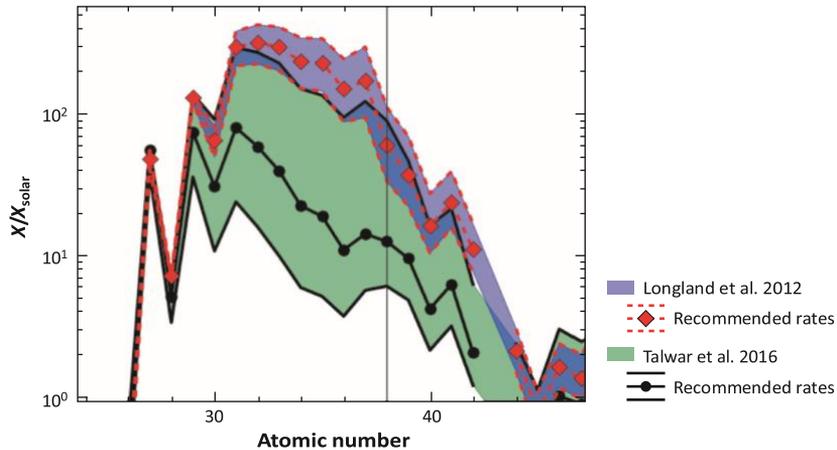

**Figure 4**

The *s*-process elemental abundance distribution relative to solar after both convective core He burning and convective shell C burning for an $M = 25M_\odot$, $Z = 0.02$ $^{22}$Ne star and its uncertainty range (based on calculations presented in Reference 99) obtained using different α-capture rates: Longland et al. (110) (*purple region* with *red lines* for the recommended rates) and Talwar et al. (99) (*green region* with *black lines* for the recommended rates). The vertical line at $Z = 38$ indicates the element strontium (Sr).

confirmed that more than 70% of Cu, Ge, and Ga in the Solar System were made by the weak *s* process (30). While production becomes less efficient for heavier elements, most of As and Rb could have been produced by this process, although this is still uncertain. From Sr and above, the *s*-process contribution from low-mass AGB stars (Section 5) becomes dominant (114, 115).

While the trend shown in **Figure 4** is generally confirmed by all stellar simulations, the *s*-process material can be partially processed and modified by the following supernova explosion depending on parameters such as the progenitor mass and the explosion energy (116, 117). Also at this stage, most of the neutrons are produced by the $^{22}$Ne(α,n)$^{25}$Mg reaction. This explosive nucleosynthesis is expected to be relevant in the determination of the isotopic abundances with at least three *s*-process branching points—at $^{63}$Ni, $^{79}$Se, and $^{85}$Kr—becoming fully activated. A detailed study of this explosive aspect of the *s* process in massive stars is still missing also because of the current nuclear uncertainties.

At low metallicity, massive stars can also produce *s*-process elements; however, they do not contribute significantly to the solar composition because the amount of the $^{22}$Ne neutron source is a direct product of the initial CNO isotopes, which decrease with decreasing initial metallicity (91, 96, 118). These stars are a marginal *s*-process source, which may affect the composition of metalpoor stars only. However, fast-rotating massive stars at low metallicity present an *s* process that is strongly enhanced compared with their nonrotating counterparts. This is due to rotationally induced mixing, which allows production of $^{22}$Ne starting from $^{12}$C made during He burning, therefore removing the dependence on the initial metallicity. The enhancement of the *s*-process abundances in fast-rotating massive stars at low metallicity was first studied by Pignatari et al. (119), who obtained the highest efficiency in the mass region between Sr and Ba. Subsequent galactic chemical evolution (GCE; see Section 7) studies showed that these stars can represent a feasible source of heavy elements in the early Galaxy (120), including contributing to the puzzling light-element primary process (LEPP) observed at low metallicity and potentially in the solar distribution (121, 122) (see Section 7). However, the latest generation of fast-rotating massive star models shows large discrepancies in the *s*-process production. In some models (123, 124),

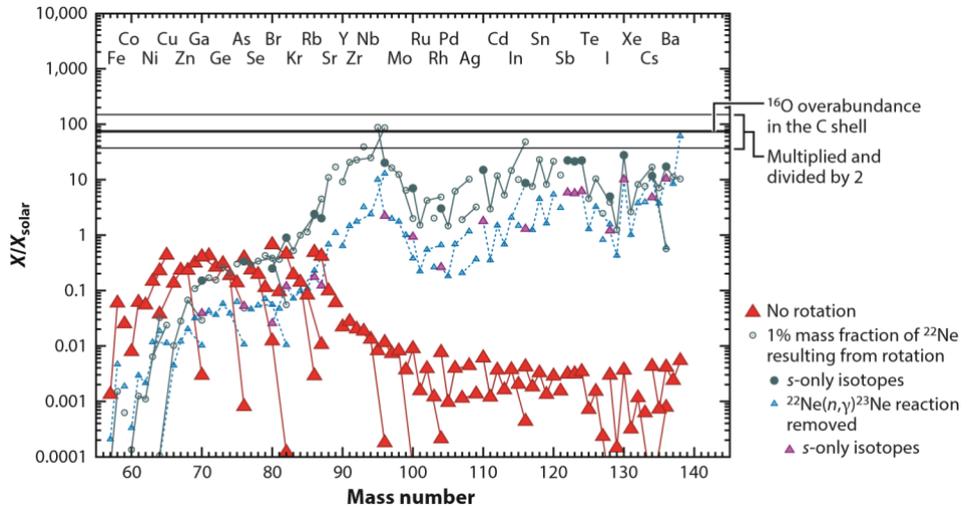

**Figure 5**

The *s*-process isotopic abundance relative to solar between $^{57}$Fe and $^{138}$Ba for a 25 $M_\odot$ star of metallicity 1/1,000 solar at the end of the convective C-burning shell, based on the models by Pignatari et al. (119). The horizontal lines correspond to the $^{16}$O overabundance in the C shell (*thick line*), multiplied and divided by two (*thin lines*). Isotopes of the same element are connected by a line, and the different symbols correspond to (*i*) no rotation (*filled large triangles*), (*ii*) a 1% mass fraction of $^{22}$Ne resulting from rotation (*open circles*; *filled circles* represent the *s*-only isotopes), and (*iii*) the same as *ii* but removing the $^{22}$Ne(n,γ)$^{23}$Ne reaction (*open small triangles*; *filled small triangles* represent the *s*-only isotopes).

*s*-process production stops at the elements in the Ba mass region, while in other models (125, 126) heavier elements up to Pb are produced. These different results have a profound impact on GCE models (127). Uncertainties in the implementation of stellar rotation in one-dimensional models are certainly the source of the variations between different models. These uncertainties need to be addressed to fully understand the role of fast-rotating massive stars in the chemical enrichment of the early Universe for light and heavy elements (120, 128, 129).

Nuclear uncertainties also have an extreme impact on the *s*-process predictions in fast-rotating stars; for example, the α-capture on $^{17}$O (97) controls the potential recycle of the neutrons captured by the neutron poison $^{16}$O. We can expect that all the rate uncertainties of relevance for the weak *s* process maximize their impact on the *s* process in fast-rotating massive stars at low metallicity. For example, **Figure 5** shows a test that highlights the role of $^{22}$Ne, a neutron source and neutron poison at the same time. If the $^{22}$Ne(n,γ)$^{23}$Ne reaction is removed from the network, the neutron exposure increases and the *s*-process production up to Ba is reduced while the abundances of $^{138}$Ba and all the following *s*-process isotopes grow by orders of magnitude.

## THE *s* PROCESS IN ASYMPTOTIC GIANT BRANCH STARS

The most prolific sources of the *s*-process elements from Sr to Pb in the Universe are stars with initial masses of roughly 1–4 $M_\odot$ during their AGB phase. These stars were shown in the 1950s to have the radioactive element Tc and enhancements in Ba and Sr. The AGB is reached after both H and He are exhausted in the core. During this phase, all the external mass is peeled off by strong stellar winds, leading to the core being left as a white dwarf (130). The structure of an AGB star is shown in **Figure 6**. The time sequence of the processes of nucleosynthesis and mixing during the AGB evolution is illustrated in the **Supplemental Video**. In brief, H and He

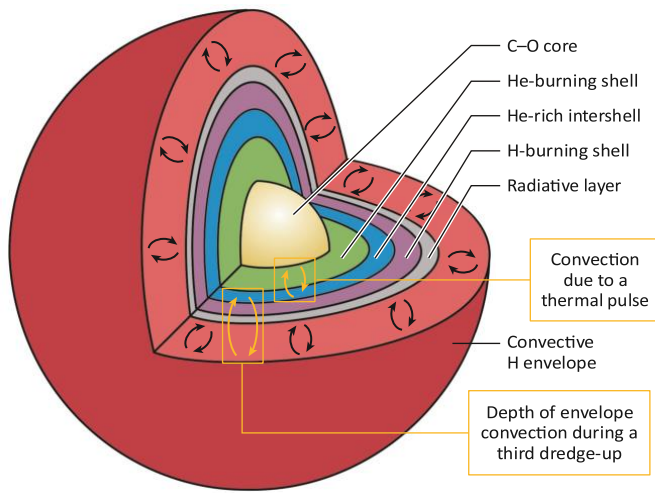

**Figure 6**

Schematic (not to scale) image of the structure of an asymptotic giant branch star. The convective H-rich envelope constitutes most of the mass of the star (e.g., for a 3 $M_\odot$ star, it is approximately 2.4 $M_\odot$) and is eroded by stellar winds, which eject material from the surface. The degenerate C-O core is the result of the H and He burning during the previous phases of the evolution. The burning shells are located between the core and the envelope. Highlighted in gold are the extent of the episodic mixing events related to the convective region generated by the thermal pulse (Phase 2; see Section 5) and to the deepening of the base of the convective envelope during the third dredge-up (Phase 4; see Section 5). The location of the deepest extent of the third dredge-up represents the point in time and space below which partial mixing is assumed to occur leading to the formation of the $^{13}$C pocket (Section 5.1). Figure provided by Magnus Vilhelm Persson (**https://figshare.com/articles/figure/Internal_structure_of_AGB_star/653683**; CC BY 4.0) and adapted (*gold parts* added) for consistency with the terminology and description in this review.

burning occur in shells around the core but not at the same time. Theoretical models predict the recurrence of a cycle where a relatively long phase of H burning ($\sim 10^3$–$10^5$ years; Phase 1) is interrupted by a burst of He burning known as a thermal pulse (TP; Phase 2). A TP lasts for $\sim 100$ years and releases a large amount of energy at the base of the He-rich intershell, which cannot be carried outward by radiation alone. Therefore, a convective region develops (Phase 3), which engulfs the whole intershell. After the convective pulse is extinguished, the envelope can penetrate the internal layers of the stars [third dredge-up (TDU); Phase 4] that were previously exposed to the convective pulse. After the mixing has reached its deepest extent, H burning starts again. The cycle (Phases 1–4) is repeated from a few to hundreds of times depending on the initial stellar mass and how long it takes for the stellar winds to eject most of the envelope.

The recurrent TDU episodes of Phase 4 carry to the stellar surface the material produced by nuclear reactions in the intershell including elements heavier than iron produced via the *s* process, together with a large amount of carbon produced by partial He burning[3] during Phases 2 and 3. Some AGB stars eventually become C-rich at the stellar surface; that is, the quantity of carbon atoms can exceed that of oxygen atoms (C > O). This has fundamental implications on the type of molecules and dust that can form in the cool ($\sim 1{,}000$ K) and dense envelopes of these stars (131).

**Intershell:** He-rich region located between the H- and He-burning shells in AGB stars

**Third dredge-up (TDU):** mixing process due to the downward movement of the convective envelope that may follow a TP, during which the base of the envelope reaches the intershell region

---

[3] In AGB stars, C is more produced than O because the triple-α reaction is activated, but the TP ends before the $^{12}$C$(\alpha,\gamma)^{16}$O reaction can also be efficiently activated.

## 5.1. Neutron Sources in AGB Stars

The *s* process in AGB stars occurs via two neutron sources, which operate at different locations and times within the intershell. The $^{22}$Ne($\alpha$,n)$^{25}$Mg reaction is activated at the relatively high temperatures (above roughly 300 MK) reached in the convective pulses of AGB stars of initial mass >3–4 $M_\odot$. The $^{13}$C($\alpha$,n)$^{16}$O reaction, activated at temperatures of roughly 90 MK, can explain the strong enhancements of neutron-capture elements observed in AGB stars of lower mass. However, there is not enough $^{13}$C in the H-burning ashes to produce the neutron flux required to match the observations. The H-burning ashes also contain large quantities of $^{14}$N, which is a neutron poison via the $^{14}$N(n,p)$^{14}$C reaction (132). A solution to this problem was found by assuming that at the deepest extent of each TDU episode, when a sharp discontinuity is present between the convective H-rich envelope and the radiative He-rich region, a partial mixing zone (PMZ) forms where protons and $^{12}$C are present together. Proton captures then lead to the formation of a so-called $^{13}$C pocket (see **Supplemental Video**). While there is no consensus on the exact mixing mechanism (Section 5.2), this model is robust enough to reproduce the observational data (Section 6.1).

There are crucial differences between the neutron fluxes from the two neutron sources. The $^{13}$C neutron source is active during the interpulse periods over long timescales (∼$10^4$ years). This results in a slow burning and a relatively low-peak neutron density (∼$10^7$ cm$^{-3}$). The $^{13}$C nuclei are usually completely burnt before the onset of the next convective pulse (25), and the neutron exposure reaches relatively high values—up to 1 mb$^{-1}$ or even higher in low-metallicity AGB stars, where the abundance of iron nuclei is lower. By contrast, the $^{22}$Ne neutron source is active for a few years only, following the sudden increase in temperature due to the TP. This results in a fast burning and a high-peak neutron density—up to ∼$10^{13}$ cm$^{-3}$ in stars more massive than 3 $M_\odot$ (48, 133). Only a few percent of the $^{22}$Ne nuclei burn, and the neutron exposure has relatively low values, typically up to 0.1 mb$^{-1}$. Overall, the $^{13}$C neutron source is responsible for producing the bulk of the *s*-process elements in AGB stars, whereas the $^{22}$Ne neutron source mainly results in abundance variations of the isotopes that are involved in the operation of branching points.

## 5.2. The Formation of the $^{13}$C Neutron Source and Effects on Its Neutron Flux

The formation of the PMZ leading to the $^{13}$C pocket has provided a good solution to the problem of the missing $^{13}$C neutron-source abundance in AGB stars (26, 134, 135). However, there is still no consensus on the physical process that drives such partial mixing, and different methods have been used by different groups in theoretical simulations. Recent models (136) have implemented a self consistent, nonparametric method based on magnetic buoyancy. While this process has also been recently included in the FRANEC stellar evolution code (137), the FRANEC models in the widely used FRUITY database (**http://fruity.oa-abruzzo.inaf.it**) (138, 139) instead include time-dependent overshoot at the base of the envelope to produce the $^{13}$C pocket. In this description, a free parameter $\beta$ scales the exponential decline of the convective velocity. To add the effect of gravity waves, the models of the NuGrid Collaboration (**https://nugrid.github.io/content/data**) include two parameter-scaled ($f_1$, $f_2$) functions at different mixing depths (140–142). In these models, the mixing is also applied to the base of the convective pulse, which results in the $^{12}$C in the intershell, and therefore the $^{13}$C and free neutrons in the pocket, being higher than in the case where convective overshoot at this boundary is not included (143).

Other models (26, 135, 144–146) have avoided defining the physical process that drives the partial mixing and have employed less self-consistent but more flexible methods. These involve directly adding into the top layers of the intershell a mixing function, a proton profile, or the final

---

**Partial mixing zone (PMZ):** region at the top of the He-rich intershell, just below the deepest extent of the mixing due to the TDU, where a proton abundance profile can form

**$^{13}$C pocket:** region of the PMZ where proton captures on 12C produce 13C, but the protons are not enough to also destroy 13C into 14N, and a pocket forms with a significant amount of 13C

**Interpulse:** the time that elapses between two TPs

**Supplemental Material >**

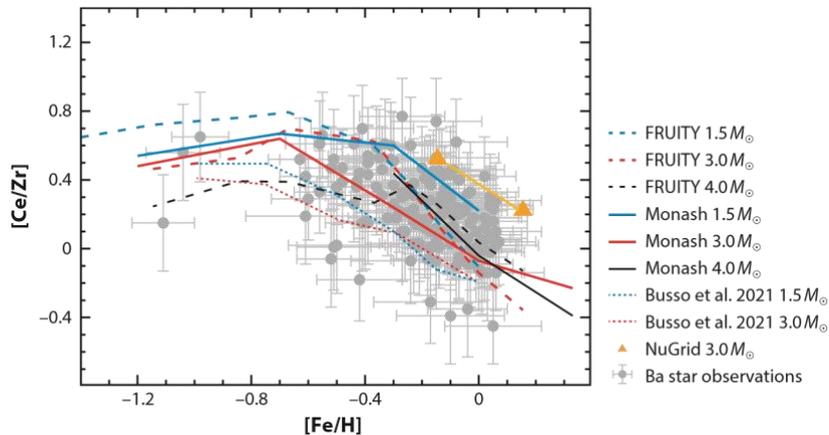

**Figure 7**

Comparison between Ba star observations of the [Ce/Zr] ratio as a function of the metallicity ([Fe/H]) and the corresponding predicted final surface ratio by asymptotic giant branch models. Models of 1.5 (*blue lines*) and 3 $M_\odot$ (*red lines*) are shown from FRUITY (139), Monash (146), and Busso et al. (137); models of 4 $M_\odot$ (*black lines*) are shown for FRUITY and Monash. Also plotted are two 3 $M_\odot$ NuGrid He07 models (*orange triangles* and *lines*) (141). The original figure from Reference 152 includes all four combinations of first-peak (Y and Zr) and second-peak (Ce and Nd) elements available from the data set (158). The results are all similar, and the [Ce/Zr] ratio is shown here because those data have the smallest error bars. Figure adapted with permission from Reference 152; copyright 2018 ESO.

$^{13}$C and $^{14}$N profiles. This approach is also supported by evidence indicating that, once some mixing is assumed to occur, the first-order effects that determine the $^{13}$C abundance profile, and therefore the neutron flux, are the rates of the reactions involved in the proton-capture chain leading to the formation of the $^{13}$C pocket. The results from the Monash models (146), which use an artificial mixing function that exponentially declines with mass (see **Supplemental Video**), are very similar to those of the standard FRUITY models that use time-dependent overshoot (see also **Figure 7**). Tests performed to check the effect of changing the shape of the mixing profile in the Monash models (147) concluded that a significant effect on the *s*-process abundances results only when the profile is modified by orders of magnitude.

The proton-capture reactions also control the main structure of the $^{13}$C pocket. Starting from the PMZ, two regions form: one at the bottom that is rich in $^{13}$C and one at the top that is rich in $^{14}$N. In the top region, the *s* process is completely inhibited by the $^{14}$N$(n,p)^{14}$C reaction (135,143). Furthermore, any process that may mix $^{14}$N into the $^{13}$C-rich region can lead to a potential decrease in the number of free neutrons. Slow mixing processes in the pocket during the interpulse may be driven by rotation (148–150), magnetic fields, gravity waves (151), and/or diffusive mixing (142). The detailed comparison of spectroscopic observations of a large sample of Ba stars with AGB model predictions not including any of these extra processes has provided general agreement (see Section 6.1) in relation to the trend of the data with metallicity (152). Therefore, these extra processes likely represent second-order effects. This conclusion is in agreement with asteroseismology observations showing that the steepness of differential rotation near the core of AGB stars cannot be as strong as what would affect the *s* process (153). In summary, while there is no consensus on the physical mixing mechanism(s) that may produce the $^{13}$C pocket and several other possible physical effects that can change the neutron flux within it, current AGB *s*-process models are robust enough to reproduce the enhancements of elements heavier than Fe observed in stars and their trend with metallicity (Section 6.1).

## 6. DIRECT OBSERVATIONAL CONSTRAINTS

### 6.1. Stellar Spectroscopy

Plenty of observational data are available for the *s* process in AGB stars thanks to medium- and high-resolution spectroscopy of stars enriched in *s*-process elements relative to solar. These stars are the AGB stars themselves (154, 155), their progeny in the form of post-AGB stars (156) and planetary nebulae (157), and stars that accreted material from the winds of a more massive binary companion during its evolution through the AGB phase. To this latter type of objects belong Ba stars, with metallicities roughly from solar to 1/10 of solar (158); CEMP stars with Ba enhancements (CEMP-*s*), with metallicities roughly below 1/100 of solar (159); and CH stars, with metallicities between those of the previous two classes (160). Such a large variety of observational data requires caution when making comparisons with AGB model predictions. For example, postAGBstars may suffer nucleosynthesis beyond the end of the AGBphase,and thus their abundances may not reflect the surface abundances predicted at end of the AGB phase. Planetary nebulae are observed in emission rather than absorption lines and provide abundances for the elements that do not condense into dust, such as Se, Kr, Te, and Xe, which are different from those that can be observed in the other objects. The origin of planetary nebulae is still not completely clarified, and it is possible that many such objects are the result of the evolution of binary systems where the AGB phase was truncated (161). Binary companions, such as Ba, CH, and CEMP stars, have the advantage of being main sequence or red giant stars hotter than AGB stars. Their atmosphere models are simpler than those of the AGB stars themselves,which are dynamical and rich in molecules and dust; thus, their modeling has large systematic uncertainties (162). Furthermore, CH and CEMP stars provide information about AGB stars [and potentially massive stars (37)] of low metallicity, which evolved in the past and are not visible anymore. The disadvantage is that the process of binary mass transfer is still not completely understood. Therefore, ratios of two *s*-process elemental abundances are the most direct constraints, as both elements in the ratio would be affected by the same dilution during the mass transfer.

A significant step forward in our understanding of the *s* process has been made thanks to highresolution observations of a self-consistently analyzed, large sample of roughly 170 Ba stars (158, 163,164).This development has helped resolve previous problems related to small-number statistics,inconsistency in the data taking and analysis of different small samples,and evaluation of error bars. A summary of the general impact of these new data on our understanding of the *s* process (152) can be derived from analysis of **Figure 7**. The [Ce/Zr] ratio [4] represents the relative production of the second to the first peak, and it increases with decreasing metallicity ([Fe/H]) in the Ba star data. This trend was expected since the 1980s, following the idea that $^{13}$C is the main neutron source in AGB stars (165). The $^{13}$C production depends not on the presence of metals but on that of the H and He (converted into $^{12}$C via the triple-α reaction) originally present in the star. The Fe abundance, the seed for the *s* process, depends instead on the initial metallicity (i.e., [Fe/H]). Decreasing the [Fe/H] results in the same amount of $^{13}$C and neutrons produced, but a lower amount of Fe capturing them. The overall result is that the number of neutrons captured by each iron seed is larger, and the *s* process can reach the second peak. Below [Fe/H] ∼−0.8, the third peak is also reached, and the [Ce/Zr] ratio becomes roughly constant because the first and second peaks are in steady-state equilibrium. Every AGB model where the $^{13}$C neutron source is efficiently activated behaves in this way; therefore, all the models shown in **Figure 7** predict the same trend as the observations.

---

[4] The spectroscopic notation used here indicates the ratio of two elements *X* and *Y*, represented as the logarithm of the ratio relative to the same ratio in the Sun: $[X/Y] = \log_{10}(X/Y)_{star} - \log_{10}(X/Y)_\odot$.

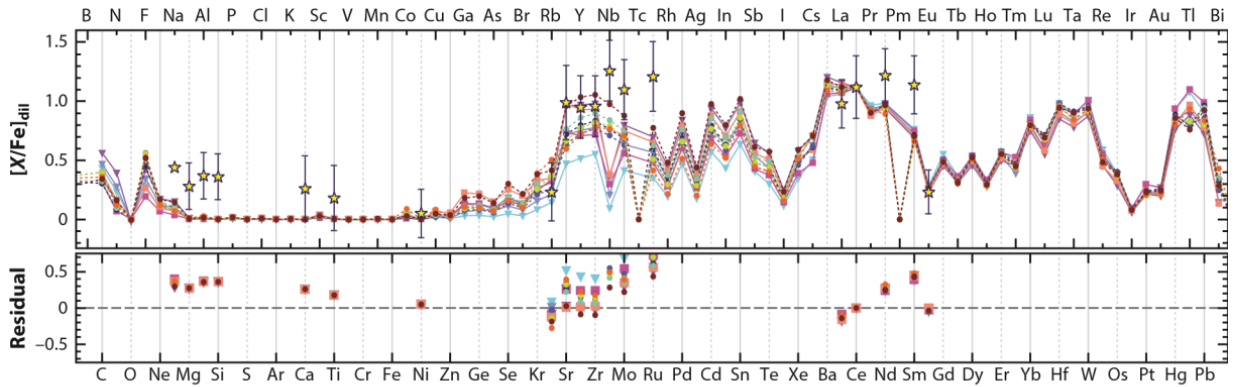

**Figure 8**

Available elemental abundances (158, 163, 164) for the Ba star CD −42°2048 (*yellow stars with error bars*), with [Fe/H] =−0.23 ± 0.16, mass $1.9^{+0.7}_{-0.5} M_\odot$, and an estimated initial mass of 3.1 $M_\odot$ for the asymptotic giant branch (AGB) companion (167), plotted with a selection of best-fit AGB models from FRUITY (*colored lines with triangles and squares*) and Monash (*colored lines with circles*) with masses in the range of 2 to 3.5 $M_\odot$ and [Fe/H] from −0.15 to −0.37. The predicted abundances are diluted such that [Ce/Fe] is exactly matched. The differences between the observations and the predictions of each model are shown in the bottom panel. Figure adapted with permission from Reference 166; copyright 2022 ESO.

Nevertheless, for each [Fe/H] there is a spread in the models, as well as in the data, of roughly a factor of two when considering the observational error bars. For example, on the one hand, the NuGrid models predict ratios higher than the other models due to overshoot at the base of the TP (as mentioned in Section 5.2). The models that include the formation of the $^{13}$C pocket via magnetic buoyancy, on the other hand, predict ratios lower than the other models because this mechanism predicts a flatter and lower $^{13}$C abundance profile. The FRUITY and Monash models also show relatively large variations depending on the initial mass, which are due to temperature effects in the activation of both the $^{13}$C and the $^{22}$Ne neutron sources. As mentioned in Section 5.2, other effects not shown in **Figure 7**, such as rotation and diffusive mixing, may also play a role, and it is difficult to disentangle which, if any, may play the major role in the observed spread. First attempts at disentangling the effect of the stellar mass are being made (166). The masses of Ba stars and their AGB companion derived from binary considerations are still relatively uncertain (167) but appear to be typically around 2–3 $M_\odot$. This mass range is consistent with the observed low Rb abundances (164), as Rb is produced only at higher masses by the operation of the $^{22}$Ne neutron source (154) and the branching point at $^{86}$Rb.

Also discovered in the sample is the presence of overabundances of the elements just beyond the first and second peaks in a fraction of Ba stars (163, 166). For example, the first and second peaks of the Ba star in **Figure 8** can be well reproduced, within the error bars, by roughly 10 different AGB models. This star has [Fe/H] ≃ −0.2 and [Ce/Zr] ≃ −0.2 and is therefore a common Ba star when considered in relation to **Figure 7**. However, it is impossible to match the observed abundances of the elements Nb,[5] Mo, Ru, Nd, and Sm because they are consistently higher than those predicted by the steady-state behavior of the *s* process between the peaks (Section 2). These elements indicate that some type of neutron-capture process beyond the *s* process is relatively common even in Ba stars. Further analysis is underway using faster and unbiased machine learning techniques together with models of the *i* process (168).

---

[5] Note that the abundance of Nb is determined by the neutron-capture cross section of the radioactive $^{93}$Zr (1.5 Myr), for which experimental data are available only on a partial energy range (169).

## 6.2. Meteoritic Materials

Another direct observational constraint for the *s* process in AGB stars comes from tiny (∼0.1– 1 μm) SiC stardust grains that have been recovered from meteorites (131, 170). Different populations of these grains have been identified, mostly depending on the isotopic composition of their main elements, C and Si. The vast majority of them (>90%) are the mainstream SiC (MS SiC) grains, which show the signature of formation in AGB stars of metallicity around solar[6] when the condition C/O > 1 is reached. This condition is required such that not all the carbon is locked up in CO molecules and some of it is available to form SiC molecules instead. MS SiC grains carry the signature of the *s* process in a large number of trace elements, such as Ni (171), Sr and Ba (172), and W (173). In fact, they were identified as the carriers of the Xe-S component—that is, Xe atoms released by meteorites upon heating showing strong excesses in the *s*-only isotopes $^{128,130}$Xe (174).

Caution is needed when interpreting the SiC data because the mass and metallicity of the AGB parent star of each grain are not known a priori. Furthermore, different laboratory techniques are applied depending on the grain size; for example, small grains (∼0.1 μm) can be analyzed only in bulk samples of millions of them, while large grains (∼1 μm) can be analyzed individually thanks to resonant ionization mass spectroscopy (RIMS) (177). In the past decades, much effort has been devoted to the interpretation of the high-precision isotopic data from single-grain RIMS measurements. For example, neutron exposures lower than those typically achieved in AGB stars of solar metallicity are necessary to match the Sr isotopic composition of these large grains, showing $^{88}$Sr/$^{86}$Sr ratios lower than solar (172). These data have been used to constrain the details of the processes discussed in Section 5.2, from the debated mechanism of the formation of the $^{13}$C pocket (178, 179) to the mixing processes within it (142). Lugaro et al. (45, 180) proposed that the Sr variation from the large to small grains (with $^{88}$Sr/$^{86}$Sr ratios varying from lower to higher than solar) is due to variation in the metallicity of the parent star—that is, to the first-order effect on *s*-process nucleosynthesis shown in **Figure 7**. Specifically, the larger grains require lower neutron exposures and originated in AGB stars of metallicity higher ($Z \sim 2Z_\odot$) than that ($Z \sim Z_\odot$) of the parent stars of the smaller grains.

Models of metallicity higher than solar also provide a best fit to the $^{92}$Zr/$^{94}$Zr ratio (**Figure 9**). At solar metallicity a match can be achieved only by specific choices of the abundance profiles in the $^{13}$C pocket (178) or by removing the overall C/O > 1 constraint for the formation of SiC via processes of magnetic tension in the stellar winds (181). **Figure 9** also reflects the fact that the branching point activated at $^{95}$Zr and producing $^{96}$Zr plays a significant role in the matching of the data. Models of *Z* higher than solar are colder than their counterparts of the same mass at solar *Z*; therefore, the $^{22}$Ne neutron source is less activated and less $^{96}$Zr is produced, contributing to the best match to the data (180, 182). An origin of large SiC grains in AGB stars of metallicity higher than solar still needs to be investigated—for example, by models of dust formation and observations of dust around AGB stars. If confirmed, it would provide evidence for the presence at the time of the formation of the Sun of the metal-rich stellar population observed today (183) in the solar neighborhood, which would have implications on its history.

Other types of meteoritic materials also carry the signature of the *s* process. Leachates are residues of sequential extraction procedures performed using acids of different strength. Stronger acids enable the hardest minerals to be dissolved, and analysis of the material coming out during

---

[6] Two other minor populations, Y and Z, which represent roughly 1% each of all stardust SiC, are also believed to have originated in AGB stars but of different masses and/or metallicities than the MS SiC grains (175, 176).

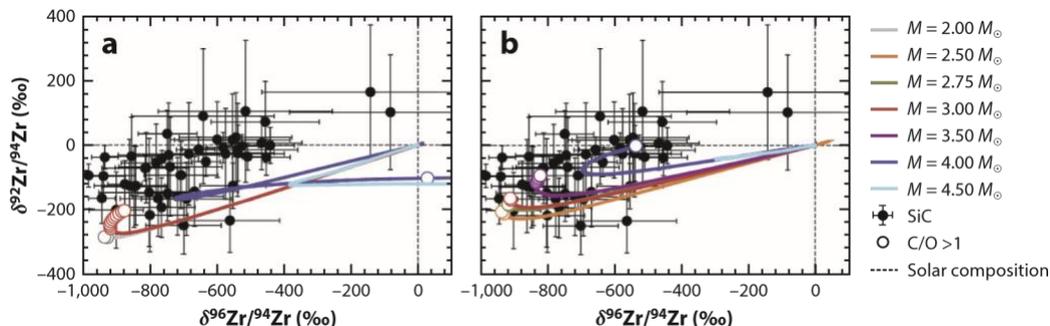

**Figure 9**

Resonant ionization mass spectroscopy silicon carbide grains data for Zr (*black circles* with 2σ error bars; for further information on the data, see Reference 178) compared with the surface evolution of Monash stellar models of (*a*) solar metallicity ($Z = 0.014$) and (*b*) twice-solar metallicity ($Z = 0.03$) of different masses from 2 to 4.5 $M_\odot$ (with colors indicated next to the plots). Open circles on the lines represent the third dredge-up episodes during which C/O > 1 in the envelope. The isotopic ratios are represented using δ values—that is, variation relative to the solar ratio, multiplied by 1,000. The dashed black lines represent the solar composition, with δ= 0 by definition. Figure adapted with permission from Reference 180; copyright 2018 Elsevier.

the process reveals the composition of the minerals that were dissolved. For example, leachates that do not include the dissolution of stardust SiC will show a deficit of *s*-process isotopes, while leachates that do will show excesses. This method is useful not only to measure *s*-process signatures but also to identify other types of stardust grains that are difficult to extract because they are, for example, too small or too easy to destroy. Leachates generally demonstrate that stardust grains with *s*-process signatures are present ubiquitously in primitive meteorites—as seen, for example, from Mo (184) and W (185) and several other elements (e.g., 186 and references therein).

The signature of the *s* process has also been found in bulk meteoritic rocks and used to identify the material that made up the terrestrial planets. In fact, the Earth appears to be more rich in *s*-process isotopes made by AGB stars, and likely carried by SiC grains, relative to any other group of meteorites representing Solar System bodies that formed both farther away and closer to the Sun (187–189). This is interpreted as evidence that *s*-process-rich material was present in the innermost part of the protoplanetary disk. Model predictions from the *s* process in AGB stars can be used to identify the stellar sources of such anomalies and therefore aid in the study of how they were distributed in the disk (188).

Finally, there are two radioactive isotopes (190, 191), $^{107}$Pd (6.3 Myr) and $^{182}$Hf (8.9 Myr), whose abundances in the early Solar System are well known from analyses of excesses of their daughter nuclei, $^{107}$Ag and $^{182}$W, in iron meteorites and calcium-aluminum-rich inclusion, respectively.[7] The abundances of these nuclei can provide constraints for the evolution of the *s* process in the Galaxy (Section 7) as well as information on the history of the Solar System matter. For example, the discovery that the branching point at $^{181}$Hf, which produces $^{182}$Hf, is activated during the *s* process in AGB stars allowed a first direct timing—relative to the formation of the Sun as provided by the meteoric data—of the last *r*- and *s*-process events that contributed to the solar abundances (77). This allowed us to determine the nature of the *r*-process event itself (192) as well as the evolution of the *s* process in the Galaxy and the lifetime of the molecular cloud where the Sun was born (193).

---

[7] Another two radioactive isotopes that can be produced by the *s* process, $^{205}$Pb (17.3 Myr) and $^{135}$Cs (2.3 Myr), have been investigated and may have been present in the early Solar System; however, the $^{205}$Pb abundance is not well determined, and for $^{135}$Cs there is only an upper limit.

# 7. THE GALACTIC CHEMICAL EVOLUTION OF THE *s*-PROCESS ELEMENTS AND ISOTOPES

The GCE of the elements heavier than iron was analyzed in detail first by Travaglio et al. (114, 121, 194), who investigated the production of the different *s*-process peaks in the Galaxy using *s*-process stellar yields from calculations of massive and AGB stars. Following those works, many different aspects of the problem have been addressed, both theoretical and observational. For example, the impact of rotating massive stars (Section 4) in the early Galaxy has been considered by various authors (127, 195) also using inhomogeneous models of GCE to study the galactic halo (120). Yields for the *r* process from model calculations, rather than parameterized, have been included in the GCE models (196, 197). Progress has been made possible also by the increase of observational data, including at low metallicity (198). One of the main topics still under discussion is the existence of the LEPP that was introduced by Travaglio et al. (121) to explain both a missing ∼20% in the predicted solar abundances of the first-peak *s*-process elements and the enrichment of these elements observed in low-*Z* stars. It is still debated whether this process is needed to produce the solar abundances (199), which sources can provide such abundances at low metallicity (122, 200–202), and whether rotating massive stars are the solution (127, 196). Analysis of the GCE of the radioactive isotopes $^{107}$Pd and $^{182}$Hf (see Section 6.2) has also indicated a problem of underproduction of $^{107}$Pd relative to $^{182}$Hf in the early Solar System. This problem may be solved if only one single last AGB star contributed to their abundances; however, since AGB stars are relatively common, this solution would require very slow mixing of AGB ejecta in the Galaxy such that only one star would have contributed the total of these two isotopes to the solar parcel of gas. The other option is to invoke an extra source of Pd in the Solar System, which may correspond to a LEPP, but at high metallicity, or may be related to other neutron-capture processes, such as those identified in Ba stars beyond the first *s*-process peak (**Figure 8**).

Recently, data on the evolution of neutron-capture elements in the Galaxy have been provided as a function of stellar age (203, 204) on top of [Fe/H]. These data make it possible to use ratios of *s*-process elements to elements made by core-collapse supernovae (e.g., [Y/Mg]) as chemical clocks—that is, indicators of stellar ages (205) as well as new complementary constraints to the GCE analyses presented so far. For example, the ratio of second- to first-peak elements is observed and predicted to decrease with increasing [Fe/H] (see, e.g., figure 17 of Reference 195, although there is a large scatter in the data), following the *s*-process trend of **Figure 7**. Simple GCE models, where the age of a star (as its birth time relative to today) decreases with increasing metallicity, therefore predict that the ratio of second- to first-peak elements should decrease with age. Instead, variable trends for different *s* elements are observed as a function of age. For example, figure 11 of Reference 204 shows a [Ba/Zr] ratio that behaves in the opposite way as predicted, and there are clear differences even for elements that belong to the same peak, such as between Sr+Y and Zr or between Ce+La and Ba, which should instead behave in the same way. Also for solar twins, which have the same metallicity of the Sun by definition, variations with stellar age have been observed (see figure 8 of Reference 203). These problems need to be investigated in relation to possible different neutron-capture processes in AGB stars of different mass (and hence different lifetimes) and the galactic processes that contribute to the observed flatness and strong scatter of the age–metallicity relationship in the Galaxy (206). Finally, strong enhancements of Ba have been observed in stars of very young ages (∼3–300 Myr) in open clusters (207). It is still unclear if they are related to observational problems in modeling the atmospheres of young stars or if they represent a galactic signature of the *i* process, which can produce Ba excesses via the accumulation of $^{135}$I (**Figure 2**).

## SUMMARY POINTS

1. Beyond the traditional s and r processes, the i process and the neutron burst (n process) located away from the valley of β stability have received much interest in the past decades due to the existence of observational constraints that cannot be resolved within the two traditional processes.

2. The neutron sources for the s and i processes as well as for the neutron burst are $^{13}$C and $^{22}$Ne nuclei. Their (α,n) reaction rates still carry uncertainties, especially in the case of $^{22}$Ne. Neutron-capture cross sections of stable and unstable isotopes as well as β decays need to be investigated for the s, i, and n processes.

3. The formation and activation of the main neutron source ($^{13}$C) in asymptotic giant branch (AGB) stars are still surrounded by many uncertainties. However, these appear to be second-order effects relative to the fact that its production does not depend on stellar metallicity.

4. Spectroscopic observations support Point 3 above. They also indicate the presence of a neutron-capture process beyond the s process in AGB stars. If data from stardust silicon carbide grains are also interpreted using Point 3, then larger grains should originate from AGB stars of metallicity higher than solar. These grains left their s-process imprint during the formation of the terrestrial planets in the Solar System.

5. The galactic chemical evolution of the s-process elements at the first peak is still a matter of debate with different models proposing extra sources beyond AGB stars. For example, fast-rotating low-Z massive stars are a possible site; however, model predictions are discordant and debated. The galactic evolution of the radioactive isotopes $^{107}$Pd and $^{182}$Hf can be reconciled with their presence in the early Solar System, although the Pd/Hf ratio is underproduced and may again be affected by neutron-capture processes beyond the standard s process occurring at solar metallicity.

## FUTURE ISSUES

1. The rates of the neutron-source reactions need to be established with error bars ∼10% at all temperatures of interest. We expect these data to be provided by ongoing underground experiments. A continuous effort is also necessary to measure neutron-capture cross sections of unstable isotopes with 10% error bars as well as their β-decay rates.

2. For both the s and i processes, the mechanism of the formation of the $^{13}$C neutron source is still debated, and for the i process, not even the astrophysical sites are known. Fully or partially parametric modeling needs to be performed for potential combinations of the s and i processes, where, for example, the i process may be activated on s-process seeds. Such models can be driven by observational constraints such as Ba stars.

3. The observations of s-process and neutron-burst signatures in meteoritic materials should be compared with predictions from recent models of AGB stars and core-collapse supernovae to fully identify the carriers of such signatures and the impact of their presence in Solar System bodies.

4. The impact of neutron-capture processes beyond the s and r processes needs to be fully investigated in models of the chemical evolution of the Galaxy; observed elemental trends with both [Fe/H] and stellar age need to be considered when comparing models and observations.

## DISCLOSURE STATEMENT

M.P. and R.R. are members of the NuGrid Collaboration (**https://nugrid.github.io/**). M.L. is a member of the LUNA Collaboration (**https://luna.lngs.infn.it/**).

## ACKNOWLEDGMENTS

M.L. and M.P. thank Evelin Bányai for help with Python and making the **Supplemental Video**. M.L. and M.P. acknowledge the support of the ERC Consolidator Grant (Hungary) program (RADIOSTAR, grant 724560), the NKFI via the (OTKA) K138031 grant, and the "Lendület2014" (LP17-2014) program of the Hungarian Academy of Sciences (Hungary). We thank the ChETEC-INFRA project funded by the European Union's Horizon 2020 research and innovation program under grant 101008324; the ChETEC COST Action (CA16117), supported by the European Cooperation in Science and Technology; the NuGrid collaboration (**http://nugridstars.org**); the Joint Institute for Nuclear Astrophysics–Center for the Evolution of the Elements (US National Science Foundation, grant PHY-1430152); and the US IReNA Accelnet network (grant OISE-1927130).M.P.thanks the Science and Technology Facilities Council (United Kingdom, University of Hull consolidated grant ST/R000840/1). R.R. acknowledges the support by the State of Hesse within the Research Cluster ELEMENTS (project 500/10.006).